\documentclass{pasj01}

\Received{$\langle$27-Apr-2018$\rangle$}
\Accepted{$\langle$19-Jun-2018$\rangle$}

\usepackage{graphicx}
\usepackage{textcomp}

\begin{document}

\title{Far-IR dust properties of highly dust obscured AGNs from the AKARI and WISE all-sky surveys}
\author{Anson Lam\altaffilmark{1}, Matthew Malkan\altaffilmark{1}, Edward Wright\altaffilmark{1}}

\altaffiltext{1}{Department of Physics and Astronomy, UCLA, Los Angeles, CA 90095-1547, USA}
\email{ansonl@astro.ucla.edu}


\maketitle

\begin{abstract}
The combination of the AKARI and WISE infrared all-sky surveys provides an unique opportunity to identify and characterize the most highly dust obscured AGNs in the universe. Dust-obscured AGNs are not easily detectable and potentially underrepresented in extragalactic surveys due to their high optical extinction, but are readily found in the WISE catalog due to their extremely red mid-IR colors. Combining these surveys with photometry from Pan-STARRS and Herschel, we use SED modeling to characterize the extinction and dust properties of these AGNs. From mid-IR WISE colors, we are able to compute bolometric corrections to AGN luminosities. Using AKARI's far-IR wavelength photometry and broadband AGN/galaxy spectral templates, we estimate AGN dust mass and temperature using simple analytic models with 3-4 parameters. Even without spectroscopic data, we can determine a number of AGN dust properties only using SED analysis. These methods, combined with the abundance of archival photometric data publically available, will be valuable for large-scale studies of dusty, IR-luminous AGNs.
\end{abstract}


\section{Introduction}
Galaxies that are optically faint but bright in the mid-IR tend to be high redshift, dusty galaxies \citep{Hwang_2013, Dey_2008, Yan_2004, Toba_2016}. Due to their optical faintness, these dust-obscured galaxies (DOGs) are easily missed by optical galactic surveys, but are readily found in IR catalogs. These galaxies show heavy extinction in their spectra, as inferred from the Balmer decrement (H$\alpha$/H$\beta$) \citep{Brand_2007}. IR luminous sources tend to dominate the bolometric luminosity function of galaxies in the local universe \citep{Soifer_1986}, and the most luminous sources are thought to be powered by active galactic nuclei (AGNs) \citep{Sanders_1988}. Understanding the physical mechanisms underlying the emission from dusty galaxies and AGNs can help to determine their role in galaxy evolution. 

Far-IR photometry is relatively sparse compared to photometry at shorter wavelengths, and it is difficult to get a complete picture of the physical properties of an AGN if we only have a portion of the full SED available. The far-IR emission from dusty galaxies can represent the bulk of the bolometric luminosity, but without far-IR data we need to rely on bolometric corrections to luminosity measurements at shorter wavelengths. \citet{Toba_2017} and \citet{Mullaney_2011} show that IR photometry can be used to reliably estimate total IR luminosity of a galaxy. However, even with monochromatic mid-IR luminosity measurements, bolometric estimates can be highly skewed due to the presence of mid-IR PAH features.

Additionally, targeted spectroscopic observations are highly time and resource intensive.  The largest spectroscopic surveys represent only a small fraction of the objects contained in photometric surveys, and analyzing these spectroscopic data sets tends to be significantly more time-consuming compared to SED analysis. Even when spectroscopy is necessary to achieve certain science goals, we need to know \textit{a priori} with a reasonably high level of confidence that our objects of interest (ex. AGNs with high dust content) are indeed the objects being targeted.

The all-sky coverage of WISE \citep{Wright_2010} and AKARI \citep{Murakami_2007} makes it possible to identify large samples of AGNs and examine their dust emission properties in the mid and far-IR \citep{Mateos_2012, Mateos_2013}. The wavelength range spanned these photometric surveys also represent the bulk of the bolometric luminosity for dusty galaxies. Additionally the deep FIR sky data from the AKARI satellite combined with the longer wavelength coverage of Herschel \citep{Pilbratt_2010} allows us to study the full IR SED of dusty AGNs. Mid-IR color selection from photometric catalogs has been shown to identify AGNs with a high degree to completeness and reliability due to emission from heated dust grains \citep{Stern_2012, Hickox_2007,Spinoglio_1989}. This is a significantly easier and more efficient method of identifying AGNs compared to using emission line diagnostics from spectroscopic data.

By using archival data to perform SED analysis, we can identify and analyze the physical properties of dust obscured AGNs without the additional effort required to perform spectroscopic analysis of these objects. With SED modeling, it is possible to quantify the degree of dust obscuration and understand the basic properties of the obscuring dust using archival data that is already publicly available. Dust modeling in the IR is complicated by the fact that dust emission is highly dependent on factors such as dust composition, emissivity, orientation, and distribution. However, we show that it is possible to describe the main IR dust components using relatively simple parametric models with a minimal number of parameters. 

In this paper, we present our results of modeling the dust properties of obscured AGNs selected from the WISE and AKARI photometric catalogs. In Section \ref{section:data}, we discuss the photometric data sets that we use and AGN selection criteria. Section \ref{section:SEDmodeling} describes the SED fitting using broadband galaxy and AGN templates from the UV to the far-IR. Section \ref{section:bolometriccorrection} discusses the distribution of the total integrated luminosity across the AGN SED and how bolometric corrections can be made to the integrated flux of shorter wavelength photometry. Section \ref{section:dust} outlines our procedure in modeling AGN dust temperatures and masses, and their relationships with different AGN properties. We summarize our conclusions in Section \ref{conclusion}. Throughout this paper we assume a $\Lambda$CDM cosmology with the parameters $\Omega_m = 0.3 $, $\Omega_{\Lambda}= 0.7$, and $H_0= 67.74 $ km/s/Mpc. An electronic table of the photometry, SED models, dust and temperature properties described in this paper is included with the online version of this paper.

\section{Data and AGN selection} \label{section:data}
\subsection{Photometric catalogs}
The WISE catalog contains photometry of over 747 million objects in 4 MIR bands (3.4, 4.6, 12, and 22 $\mu$m), with respective detection limits of 0.08, 0.11, 1, and 6 mJy in each band. We identify AGN candidates from the AllWISE data release \citep{Mainzer_2011,Wright_2010} by applying a color cut across the first two WISE bands such that $W1-W2 \geq 0.8$. This color selection criterion has been shown to select mid-IR AGN candidates to a high level of completeness and reliability \citep{Stern_2012}, and includes both unobscured (type 1) and obscured (type 2) AGNs. WISE objects that satisfy $W1-W2 \geq 0.8$ which have a sky density of $61.9\pm5.4$ AGN candidates per square degree to a depth of $W2 \sim 15.0$ (the AllWISE catalog has a density of $\sim 18,000$ objects per square degree, averaged across the entire sky). The AllWISE catalog also contains entries from the 2MASS extended source catalog \citep{Jarrett_2000,Skrutskie_2006}, which provides $J$, $H$, and $K$ band photometry when available.

\begin{figure}
\begin{center}
\includegraphics[width=1.0\columnwidth]{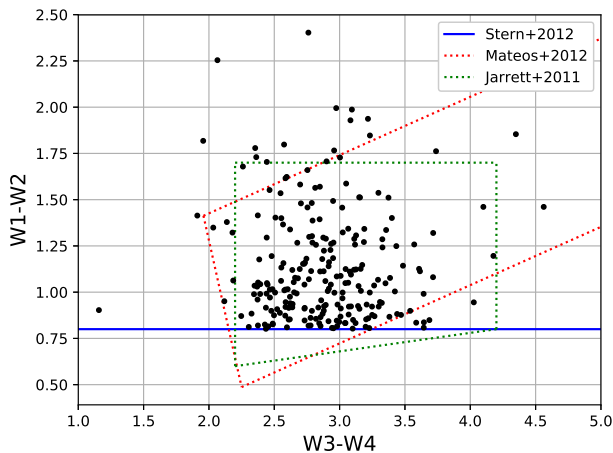}
\caption{{\label{fig:sample_wisecolors} $W1-W2$ and $W3-W4$ color diagram of the AGNs selected for SED analysis. Dotted lines show the WISE AGN selection criteria given by \citet{Stern_2012}, \citet{Mateos_2012}, and \citet{Jarrett_2011}
}}
\end{center}
\end{figure}

There are a number of AGN selection criteria available in the literature, but we choose to use the \citet{Stern_2012} color cut because it is less restrictive in the WISE color space and only uses the $W1$ and $W2$ bands. 83\% of the objects in the WISE catalog contain a detection greater than $2\sigma$ in both the $W1$ and $W2$ bands (which are more sensitive compared to $W3$ and $W4$), while only 3.5\% have detections in all four WISE bands. The color criteria described in \citet{Jarrett_2011} and \citet{Mateos_2012} use both $W1-W2$ and $W3-W4$ colors, and this severely limits the number of potential AGNs that can be identified. In Figure \ref{fig:sample_wisecolors}, we show the sample of AGNs that we use for SED fitting plotted in the $W1-W2$ and $W3-W4$ color space and a comparison of different WISE AGN color selection criteria. Since our sample has heavy Malmquist bias and contains particularly bright objects (see Section \ref{section:bolometriccorrection}), we are able to get detections in all four WISE bands. The majority of our AGNs fall within the common overlap of all these criteria, so the choice of color selection criteria does not substantially affect our sample.

The AKARI/FIS all-sky survey Bright Source Catalog \citep{Kawada_2007} contains far-IR photometry (65, 90, 140, and 160 $\mu$m) of $\sim 427,000$ objects, with respective $5\sigma$ sensitivities of 2.4, 0.55, 1.4, 6.3 Jy in each band. We cross-correlate the AKARI Bright Source Catalog with the WISE catalog using a 20'' matching radius (determined by the 40'' PSF of the AKARI 90 $\mu$m data), which yields 32,523 matches. Out of these matches, there are 2,700/3,006/9,053/2,769 detections in the 65/90/140/160 $\mu$m bands, respectively. Out of these sources, 2,609 (96.6\%)/29,577 (98.2\%)/8,722 (96.3\%)/2,630 (94.9\%) objects have reliable photometry (i.e. a flux density quality flag of 3 in the catalog) in the 65/90/140/160 $\mu$m bands, respectively. 2551 (7.8\%) of the matched WISE-AKARI sources satisfy the $W1-W2>0.8$ color cut. 

\begin{figure}
\begin{center}
\includegraphics[width=1.0\columnwidth]{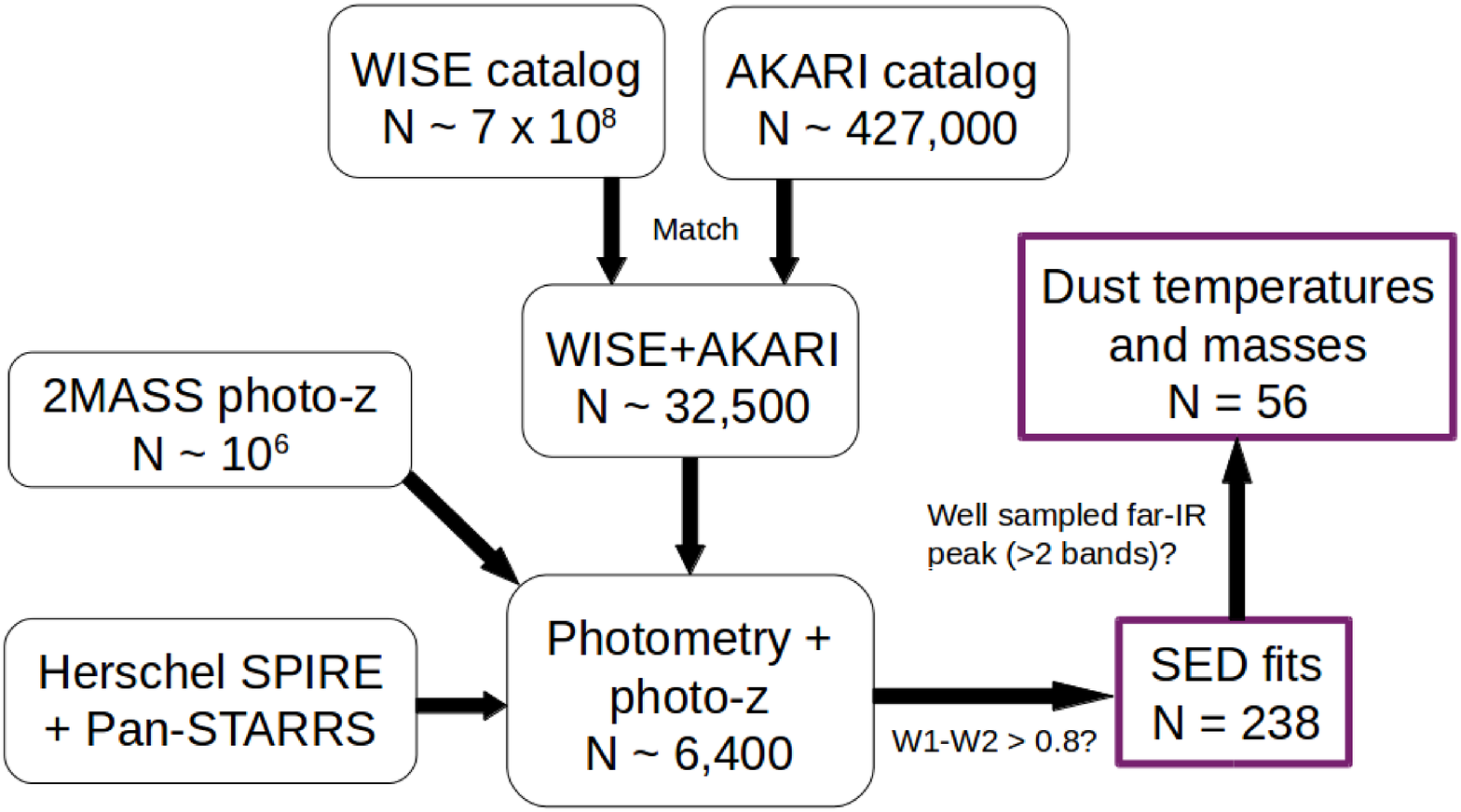}
\caption{{\label{fig:selection_flowchart} Flowchart outlining the AGN selection criteria and number of objects at each step. The purple boxes indicate the final two samples that we use for SED fitting and dust temperature/mass fitting. 
}}
\end{center}
\end{figure}

To perform broadband SED fitting, we add additional optical photometry from Pan-STARRS and far-IR photometry from Herschel to the objects in our sample when available. The SPIRE sub-millimeter camera \citep{Griffin_2010} on Herschel provides 250, 350, and 500 $\mu$m photometry. The majority of the objects in the WISE-AKARI sample do not have SPIRE detections in the FIR, perhaps as a result of the SEDs being dominated by a AGN dust component that peaks in the MIR bands instead of the FIR \citep{Hwang_2013}. The IRAS catalog \citep{Neugebauer_1984} is often used for IR studies due to its all-sky coverage, but we find that this data does not improve our sample size of objects. The WISE $W3$ and $W4$ bands are much more sensitive than the corresponding IRAS bands and contain roughly twice as many detections. Among the catalog objects with photometric redshifts (see Section  \ref{section:photoz}), AKARI has roughly twice as many detections in the 65 and 90 $\mu$m bands compared to the IRAS 60 and 100 $\mu$m bands. For these reasons, we have not included IRAS data in our analysis. For optical photometry we use the Pan-STARRS $3\pi$ Steradian Survey \citep{Chambers_2016}, which covers the $grizy$ bands. Pan-STARRS is not an all-sky survey like WISE or AKARI, but we find there are 50-60\% (depending on the Pan-STARRS filter) of the $\sim 32,000$ WISE+AKARI objects included in the catalog. In order to perform optical extinction measurements, we only include objects that contain Pan-STARRS photometry in our sample. Figure \ref{fig:selection_flowchart} show a flowchart summarizing our selection process and final number counts.

\begin{figure}
\begin{center}
\includegraphics[width=1.0\columnwidth]{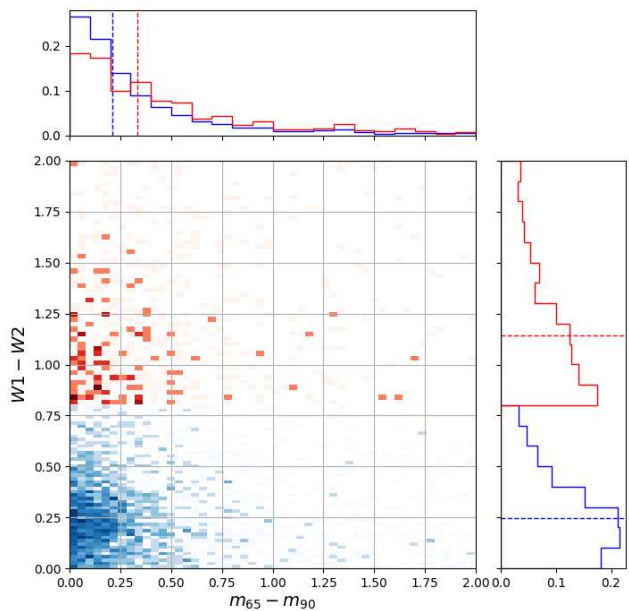}
\caption{{\label{fig:w1w2-m65m90} WISE-AKARI color space diagram, with $W1-W2$ color plotted against $m_{65}-m_{90}$ color. Objects are divided into redder ($W1-W2>0.8$) and bluer ($W1-W2<0.8$) samples based on our AGN color selection criteria. Top and right panels show histograms of relative frequencies, with dashed lines indicating the medians of the distributions.
}}
\end{center}
\end{figure}

\begin{figure}
\begin{center}
\includegraphics[width=1.0\columnwidth]{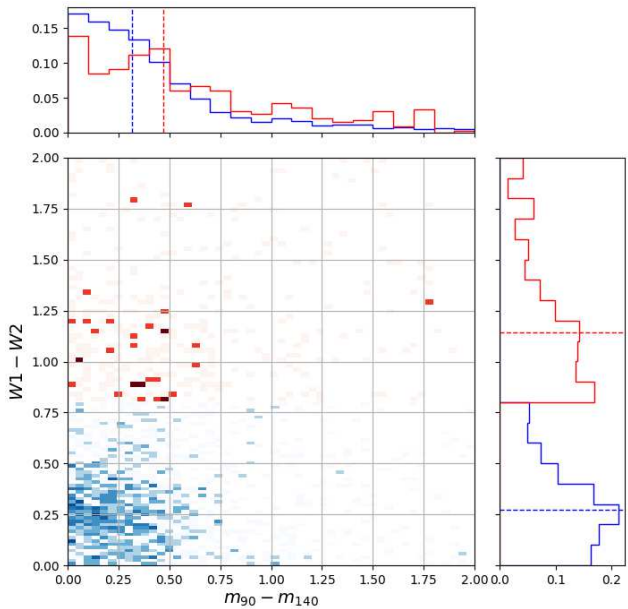}
\caption{{\label{fig:w1w2-m90m140} WISE-AKARI color space diagram, with $W1-W2$ color plotted against $m_{90}-m_{140}$ color. Top and right panels show histograms of relative frequencies, with dashed lines indicating the medians of the distributions.
}}
\end{center}
\end{figure}

Figures \ref{fig:w1w2-m65m90} and \ref{fig:w1w2-m90m140} show color diagrams of the objects in WISE-AKARI color space. The redder WISE selected sample tend to have redder $m_{65}-m_{90}$ and $m_{90}-m_{140}$ on average, which is probably a result of band shifting effects at higher redshifts. There does not appear to be any notable correlation between the WISE and AKARI colors for individual objects. 

\subsection{Photometric redshifts} \label{section:photoz}
\begin{figure}
\begin{center}
\includegraphics[width=1.0\columnwidth]{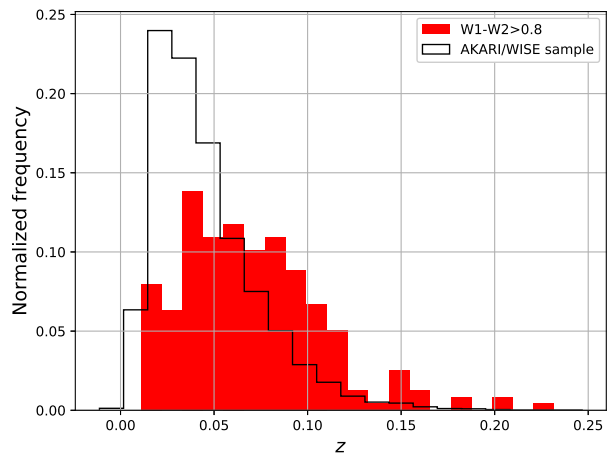}
\caption{{\label{fig:redshifts} Redshift distribution of the objects in the AKARI/WISE catalog (black curve) that have corresponding photometric redshifts in the 2MASS Photometric Redshift catalog, and the objects from that sample that satisfy the $W1-W2>0.8$ color cut (red).
}}
\end{center}
\end{figure}

In order to identify the largest possible sample of AGNs for SED analysis, we rely on photometric redshifts derived from all sky surveys. AGNs with high levels of optical obscuration are easily missed by large scale surveys such as SDSS, so there is not always spectroscopic redshift data available. We use the 2MASS Photometric Redshift catalog \citep{Bilicki_2013}, which contains a million galaxies with an accuracy of $\sigma_z=0.015$ and a median redshift of $z\sim 0.1$. These $z_{photo}$ values are derived using an artificial neural network applied to photometry in the 2MASS, WISE and SuperCOSMOS catalogs. Matching this redshift catalog with the AKARI-WISE catalog described above produces 6,384 matches. 238 of these objects meet the $W1-W2>0.8$ color cut. Figure \ref{fig:redshifts} shows the redshift distribution of these samples. 

Photometric redshifts are difficult to determine for AGNs due to their strong power law continuum and general lack of prominent emission features in the SED. This is especially true for particularly luminous AGNs, in which most of the mid-IR PAH features are absent due to photodisassociation by the AGN emission. Even with WISE photometry, the width of the $W1$ and $W2$ bands and large gaps between the $W2$, $W3$, and $W4$ bands do not allow emission features to be easily visible (in contrast, \citet{Goto_2015} exploits AKARI's near-continuous filter coverage in the mid-IR for $z_{photo}$ measurements in AKARI NEP field). Although it is possible to use the photometry that we have to estimate photometric redshifts using codes such EAZY \citep{Brammer_2008} or LEPHARE \citep{Arnouts_1999,Ilbert_2006}, the strong AGN power-law behavior makes it extremely difficult to reliably estimate photometric redshifts for these objects without additional photometry. To ensure that the redshift estimates that we use are consistent and to avoid introducing additional systematic errors, we do not perform additional photometric redshift fitting for the AKARI objects that lack $z_{photo}$ values in the 2MASS Photometric Redshift catalog.

\begin{figure}
\begin{center}
\includegraphics[width=0.7\columnwidth]{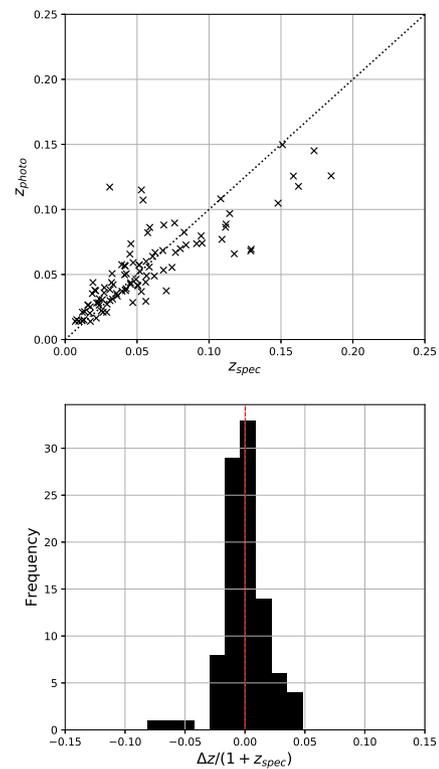}
\caption{{\label{fig:zspec-zphot} [Top] A comparison of $z_{photo}$ vs $z_{spec}$ values from the 2MASS Photometric Redshift catalog of AGNs we've selected for SED analysis. Dotted line denotes where  $z_{photo} =z_{spec}$. [Bottom] Histogram of the fractional redshift errors $\Delta z/(1+z)$. Red dashed line indicates the mean.

}}
\end{center}
\end{figure}

The 2MASS Photometric Redshift catalog also contains spectroscopic redshift values for a subset of its objects. 62\% of the $\sim 6400$ matched WISE/AKARI objects with photometric redshifts have $z_{spec}$ values. Figure \ref{fig:zspec-zphot} shows a comparison of $z_{spec}$ and $z_{photo}$ for the sample of AGNs we’ve selected for SED analysis. 97 out of these 248 AGNs have $z_{spec}$ values in the catalog. The bottom panel of Figure \ref{fig:zspec-zphot} shows that the fractional errors $\Delta z/(1+z)$ are on the order of a few percent. \cite{Ilbert_2006} defines the photometric redshift error as $\sigma_{\Delta z/(1+z)} = 1.48 \times \textrm{median} [|\Delta z|/(1+z)]$. For our sample of AGNs, we find $\sigma_{\Delta z/(1+z)}=0.013$, which is consistent with the $\sigma_z=0.015$ value given by \citet{Bilicki_2013}. 

\begin{figure}
\begin{center}
\includegraphics[width=0.7\columnwidth]{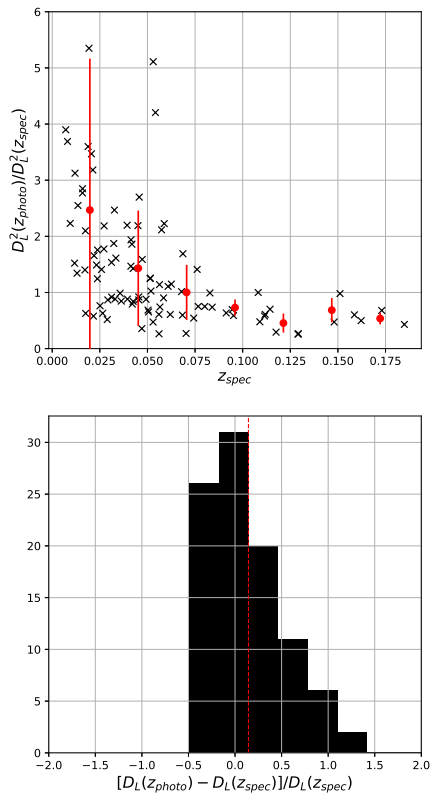}
\caption{{\label{fig:dLcomparison} [Top] $D_{L}^2(z_{photo})/D_{L}^2(z_{spec})$ plotted against spectroscopic redshift for our sample of AGNs selected for SED analysis. Red points show binned averages and error bars. [Bottom] Histogram of the fractional errors on $D_{L}(z_{spec})$, given by $[D_{L}(z_{photo})-D_{L}(z_{spec})]/D_{L}(z_{spec})$. Red dashed line indicates the mean.
}}
\end{center}
\end{figure}

One concern for our low redshift sample is that for a given error $\sigma_z$, this produces larger errors in the luminosity distance $D_L$ at low $z$ compared to higher $z$. Bolometric luminosity scales with the luminosity distance as $L \propto D_L^2$, so the relative difference in luminosities that result from photometric and spectroscopic redshifts is given by $L_{bol,photo}/L_{bol,spec} \propto D_{L}^2(z_{photo})/D_{L}^2(z_{spec})$. Figure \ref{fig:dLcomparison} shows a plot of $D_{L}^2(z_{photo})/D_{L}^2(z_{spec})$  as a function of $z_{spec}$. For the lowest redshift objects at $z\sim 0.01$, $z_{photo}$ tends to overestimate the luminosity by a factor of a few, but this factor decreases with increasing redshift. At the higher end of our redshift distribution $z_{photo}$ underestimates the luminosity by a factor of $\sim 2$. These discrepancies are within an order of magnitude, so this should not have a significant impact on our results given that $L_{bol}$ for our sample tends to span a few orders of magnitude (see Section \ref{section:IRluminosity} for our results that involve IR and bolometric luminosity estimates). The bottom panel of Figure \ref{fig:dLcomparison} shows a histogram of the fractional errors $[D_{L}(z_{photo})-D_{L}(z_{spec})]/D_{L}(z_{spec})$, and we can see that in general $z_{photo}$ tends to overestimate the luminosity distance for our sample of objects.

There are other catalogs of photometric redshifts in the literature that are also potentially useful for photometric studies of AGNs. \citet{DiPompeo_2015} provides a catalog of quasar identifications and redshifts generated using probability density functions of over 5 million objects using a combination of WISE and SDSS photometry. The authors show that their probabilistic quasar selection is more complete than the \citet{Stern_2012} WISE color-cut criteria, which tends to miss quasars that are at redshifts of $z>3$. This is because high redshift objects tend to be too blue in $W1-W2$ to satisfy the color cut. Using a 20'' matching radius to match this catalog with the AKARI BSC produces 5,079 matches, which is although smaller than the number of matches with the \citet{Bilicki_2013} catalog, represents a different set of objects located at a larger range redshifts. However, this catalog is probabilistic in nature and includes objects that perhaps have only a very small AGN component or multiple peaks in the redshift probability distribution. Since we are interested in examining the impact of AGN dominated galaxies on their surrounding dust to ensure consistency in photometric redshift measurements, we choose to limit our sample to the objects contained in \citet{Bilicki_2013}.

\section{SED modeling} \label{section:SEDmodeling}
\subsection{Optical to mid-IR SEDs ($\lambda<10\mu$m)}
\begin{figure}
\begin{center}
\includegraphics[width=1.0\columnwidth]{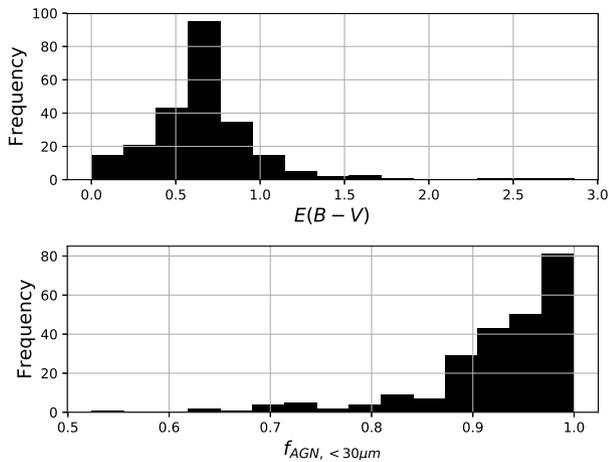}
\caption{{\label{fig:ebv-extinction} [Top] Distribution of optical $E(B-V)$ extinction values derived from SED fitting using the templates of \citet{Assef_2010}. [Bottom] Distribution of the fractional AGN contribution to the total integrated luminosity for $\lambda<30\mu$m.
}}
\end{center}
\end{figure}

We use the low-resolution spectral templates presented in \citet{Assef_2010} and model each quasar as a linear combination of an AGN template and a single galaxy template (spiral, elliptical or irregular) similar to the method described in \citet{Hainline_2014}. We find the combination that provides the best least-squares fit, and apply the extinction curves described in \citet{Gordon_1998} and \citet{Cardelli_1989} to the AGN template to simulate the effects of optical obscuration. To ensure that our photometry is reliable enough for SED modeling, we only use WISE photometry with a signal-to-noise ratio $\geq 3$, and AKARI measurements that are marked as ``high'' quality. The templates provided in \citet{Assef_2010} only extend to rest frame 30 $\mu$m, but this still encompasses all the WISE bands for the AGNs in our sample since these objects are low redshift ($z\sim 0.1$). The top panel of Figure \ref{fig:ebv-extinction} shows a histogram of the $E(B-V)$ extinction measured from these SED fits, which we use to quantify the degree of optical obscuration. We define the AGN fraction to be $f_{AGN,<30\mu m} = F_{AGN,<30\mu m}/F_{total,<30\mu m}$, which covers the wavelength range spanned by the \citet{Assef_2010} templates. The bottom panel of Figure \ref{fig:ebv-extinction} shows that the majority of the objects in our sample also have $>90\%$ of their flux for $\lambda < 30\mu$m coming from the AGN component, which gives us confidence that most of these objects are indeed AGNs. \citet{Dey_2008} shows that there is a distinction between ``bump'' and ``power-law'' type DOGs, with the latter having an AGN dust component at mid-IR wavelengths. From visually inspecting the SEDs of our sample galaxies, all of our objects show strong power law behavior in the mid-IR bands as captured by WISE. However, the large wavelength gaps between each band (particularly $W2$, $W3$, and $W4$) may not capture any PAH emission that is present. Even so, there is no indication that the mid-IR portion of the SEDs have significant curvature.

\begin{figure}
\begin{center}
\includegraphics[width=1.0\columnwidth]{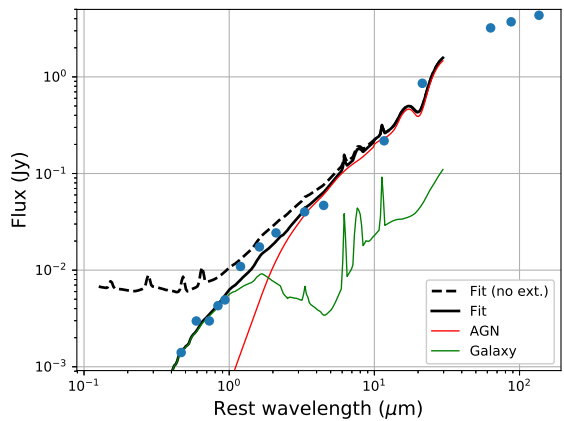}
\caption{{\label{fig:sed-assef} SED fit for using the templates from \citet{Assef_2010} (red and green curves), which extend up to 30 $\mu$m.. The solid curve shows the sum of the templates with extinction applied, while the dashed curve shows the model without extinction. 
}}
\end{center}
\end{figure}

\subsection{Mid to far-IR SEDs ($\lambda>10\mu$m)}
\label{section:ir-sed}
To examine the mid to far-IR properties of these AGNs, we use the the AGN and galaxy templates in \citet{Mullaney_2011}, which are produced from Spitzer spectra and IRAS photometric data. We once again fit each set of photometric band with a linear combination of a AGN (out of a possible set of three templates:'Mean', 'LowLum', 'HighLum') and host galaxy template (out of a possible set of six spiral galaxy templates, ranging from SB1 to SB5). Galaxies that lack far-IR SPIRE photometry (i.e. only containing photometry up to rest frame  $\sim 100 \mu$m) tend to be fit with an SED dominated by a host galaxy component. As a result, our least-squares fitting routine tends to favor the host galaxy template if there isn't a sufficient number of far-IR bands that captures both sides of the IR peak. 

Since the shape of the far-IR peak for most objects can be reasonably well fit using either the AGN or host galaxy template, it is difficult to determine what is the dominant source of dust heating in the far-IR using this kind of SED fitting alone. Since we are fitting the mid and far-IR portions of the SED independently of each other, a number of our objects are fitted with objects that are inconsistent between the two halves of the SEDs (ex. an AGN dominated template below 30 $\mu$m, and a spiral galaxy dominated template above 30 $\mu$m). There is some ambiguity in choosing the appropriate template in the far-IR portion of the SED, so we choose the type of template (AGN or starburst) to match the dominant template in the mid-IR portion. The majority of our objects already have an AGN fraction greater than 90\% in the $<30 \mu$m range, so including the far-IR portion of the SED in computing AGN fraction will likely not drastically change these values. Dusty AGNs tend to have minimal extinction in the mid-IR compared to the optical, so we do not apply any reddening correction in that portion of the SED.

\section{IR bolometric correction} \label{section:bolometriccorrection}
Since the majority of objects in the AKARI/WISE catalog do not have far-IR photometry extending beyond the AKARI bands (ex. from Herschel SPIRE), we are unable to measure the bolometric luminosity of the AGNs by directly integrating over the entire wavelength range. This can be problematic, since a number of physical parameters  (ex. Eddington ratio) that characterize an AGN depend on having an accurate estimate of $F_{bol}$. Even with  far-IR data, bolometric corrections to far-IR photometry are not necessarily reliable. For example, \citet{Elbaz_2010} shows that extrapolations from the Herschel SPIRE bands tends to overestimate the total IR luminosity, presumably because of the lack of spectral coverage in templates at those wavelengths. Mid-IR photometry is also occasionally used to estimate $L_{bol}$, but highly skewed estimates result if the band in question is centered on the 7.7 $\mu$m PAH complex. Even so, mid-IR PAH features tend to be stronger in star-forming galaxies, whereas AGNs tend to be dominated by a strong continuum. To avoid this biasing, it is preferable to estimate the total luminosity using wider wavelength range of SED photometry.

Since the WISE and AKARI bands capture most of the bolometric luminosity, the bolometric correction applied to this wavelength range should be relatively small. In general, there is more photometric data available at shorter wavelengths since this is easier to obtain. The issue then is how to apply bolometric corrections to the integrated flux from the shorter wavelength portion of SEDs.

Bolometric corrections become more important for dusty galaxies, since increasing dust content generally tends to increase the far-IR luminosity. This is easily seen in the mid-IR, in which the power law continuum tends to steepen as the far-IR dust luminosity increases. By combining the templates of \citet{Assef_2010} and \citet{Mullaney_2011}, we can see how the bolometric correction changes as a function of dust content. We use the mid-IR WISE colors to parameterize the hot dust content, since this is an observable that is generally independent of the details of any dust model. 

\begin{figure}
\begin{center}
\includegraphics[width=1.0\columnwidth]{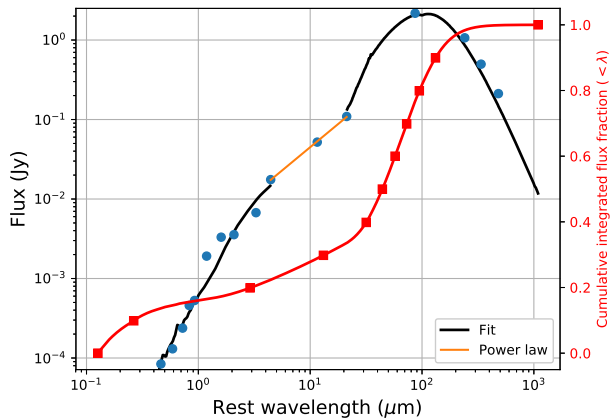}
\caption{{\label{fig:sed-cdf} Example of a broadband SED fit. The black curve shows the best fit SED, and the yellow curve is the power law continuum that we fit between the $W2$ and $W4$ bands. The red curve shows the cumulative integrated flux (ranging from the lower limit of the SED, up to the corresponding wavelength) as a fraction of $F_{bol}$. Red squares denotes increments of 0.1 along the vertical axis (flux fraction).
}}
\end{center}
\end{figure}

The SED templates from \citet{Assef_2010} and \citet{Mullaney_2011} that we are using do not overlap perfectly well in the mid-IR, since these are empirically derived from different data sets. To reconcile the two using the WISE photometry that we have, we join the templates using a power law continuum by doing a fit between the $W2$ and $W4$ bands, and integrate over the entire SED to get $F_{bol}$. This power law interpolation of course leaves out any PAH emission, but the contribution of PAH emission features to $F_{bol}$ is on the order of a few percent, with some of that emission partially canceled out by the PAH absorption at 9.7 $\mu$m. An examination of the $\nu F_{nu}$ vs. $\nu$ plots of these SEDs shows that the integrated flux converges within the wavelength range spanned by the templates. Figure \ref{fig:sed-cdf} shows an example of an SED with an power law mid-IR continuum, and the cumulative integrated flux (expressed as a fraction of $F_{bol}$) plotted against wavelength. For most of these SEDs, the cumulative flux fraction increases sharply between $\sim 10$ $\mu$m up to the far-IR peak at $\sim 100$-$200$ $\mu$m, and this region encompasses roughly $\sim 70$-$80\%$ of the bolometric flux.

\begin{figure}
\begin{center}
\includegraphics[width=1.0\columnwidth]{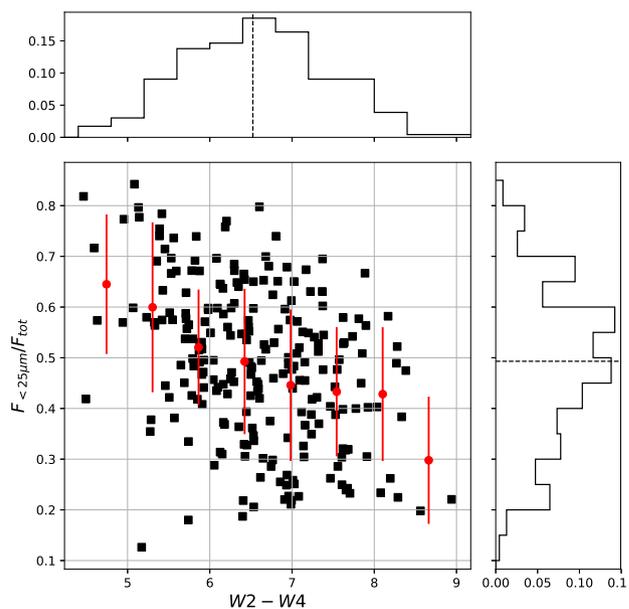}
\caption{{\label{fig:f25-w2w4} The fraction of the total luminosity below 25 $\mu$m  plotted against $W2-W4$ color. Red points show binned averages and error bars.
}}
\end{center}
\end{figure}

Since WISE photometry is generally more abundant compared to far-IR photometry and captures a significant fraction of the bolometric luminosity, this is useful for predicting $F_{bol}$. Figure \ref{fig:f25-w2w4} shows the fraction of the bolometric flux below 25 $\mu$m (roughly the upper limit of WISE) plotted against $W2-W4$ color for our AGN sample. From the binned averages, there is a clear trend of very red objects in the mid-IR having a smaller percentage of $F_{bol}$ under 25 $\mu$m, with the median being $F_{<25\mu m}/F_{bol} \sim 0.5$. This is consistent with our expectation that a larger fraction of the bolometric luminosity should be found at longer wavelengths as the slope of the mid-IR continuum increases. In other words, very red, dusty objects are dominated by far-IR emission in their energy budget. Although the scatter is considerably large, it is still possible to make a rough bolometric flux correction using this relation. Since our sample is very heavily Malmquist-biased towards highly far-IR luminous AGNs, the points in Figure \ref{fig:f25-w2w4} probably represent upper limits to the bolometric correction. 

\begin{figure}
\begin{center}
\includegraphics[width=1.0\columnwidth]{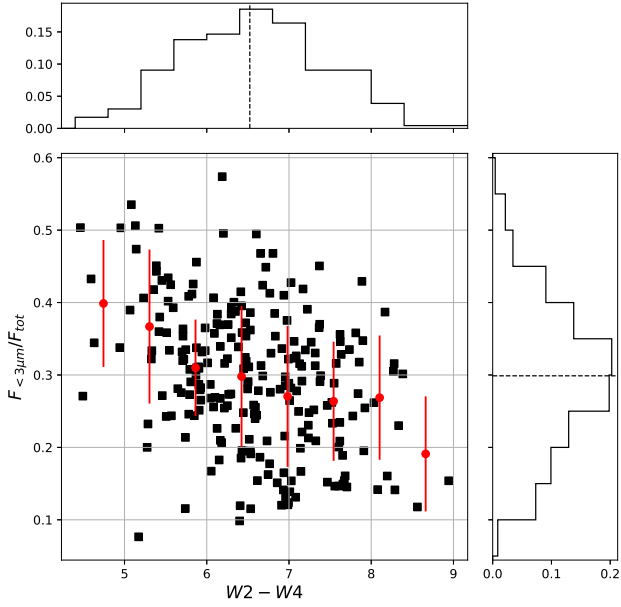}
\caption{{\label{fig:f3-w2w4} The fraction of the total luminosity below 3 $\mu$m  plotted against $W2-W4$ color. Red points show binned averages and error bars.
}}
\end{center}
\end{figure}

Since the shapes of the optical and near-IR portions of the dusty AGN SEDs tend to be highly variable across objects due to different amounts of extinction, it is difficult to produce a general template in that regime that can be applied to a variety of AGNs. However, the $\nu F_{\nu}$ contribution of the shorter wavelength region tends to be substantially smaller than the mid-IR. In Figure \ref{fig:f3-w2w4} we show the fraction of the bolometric flux below 3 $\mu$m (the lower limit of WISE) plotted against $W2-W4$ color. The change in $F_{<3\mu m}/F_{bol}$ with $W2-W4$ is smaller than the previous plot, with the median being $F_{<3\mu m}/F_{bol} \sim 0.3$. 

Since the optical and UV flux that is absorbed by dust should be re-processed and re-emitted in the IR, one might expect some sort of relationship between the level of extinction and the IR dust luminosity. This requires us to make some assumptions in the physical processes of optical dust extinction, and one of the biggest assumptions is that the level of extinction (ex. $E(B-V)$) scales directly with the quantity of dust. However, we observe no apparent correlation between $E(B-V)$ and the mid or far-IR luminosity within any wavelength range. This suggests that other factors (ex. dust distribution and geometry) likely play a role in the extent of optical dust absorption. In general, dusty AGNs tend to have highly variable optical/UV SEDs, but the general shape of the mid and far-IR SED remains does not change significantly between objects.

\section{Dust temperature and mass} \label{section:dust}
\subsection{Modified blackbody and power law fitting}
There are a number of SED template models in the literature that are given as function of $L_{IR}$ and dust temperature \citep{Chary_2001, Dale_2002}. However, dust temperatures corresponding to SED templates are necessarily discretized, and using galaxy template sets offers less flexibility than analytic models in choosing and constraining model parameters. For functional fits, a number of papers in the literature assume an isothermal dust model and fit the far-IR portion of the SED with a modified blackbody $S(\nu) \propto (1-e^{-\tau(\nu)})B_{\nu}(T)$, where $(1-e^{-\tau(\nu)})$ term accounts for dust opacity and emission. The optical depth $\tau$ is parameterized by $\tau(\nu) = (\nu/\nu_0)^{\beta}$, where $\nu_0$ is the frequency at which $\tau=1$ \citep{Draine_2006}. In the optically thin case $\tau \ll 1$, this reduces to $1-e^{\tau} \sim \nu^{\beta}$. \citet{Amblard_2010} represents the modified blackbody with emissivity function $\epsilon_{\nu} \propto \nu^{\beta}$ :
\begin{equation} \label{eq:powerlaw-BB}
S_{\nu} = \epsilon B_{\nu} \propto \frac{\nu^{3+\beta}}{e^{h\nu/kT_0} -1}.
\end{equation}

There are several issues with using a single temperature model, particularly if the SED is well sampled in the far-IR. A single modified black body function captures the far-IR peak emission, but tends to systematically underpredict the observed fluxes at wavelengths longer than the peak. \citet{Dunne_2001} accounts for this by instead using a two temperature dust model with an additional low temperature dust component (typically $T\sim 15$K). 

There is also emission in the mid-IR that is not accounted for by a single or two temperature model \citep{Casey_2012}. This mid-IR excess is due to hotter dust components (with several temperature components) that is heated by a central AGN. This requires additional blackbody components to be added, usually three or more to adequately describe the smooth power law nature of the mid-IR emission \citep{da_Cunha_2008}. Thus, an accurate fit of IR dust emission requires an arbitrary number of blackbody components, each with their own set of parameters. Fitting such a large set of parameters becomes unwieldy, especially if there are only a limited number of data points available. It is possible in principle to represent the dust emission as a temperature distribution convoluted with a modified blackbody function, but this is a computationally expensive procedure. 

As a compromise, we use a single modified blackbody with a mid-IR power law component as described in \citet{Casey_2012}:
\begin{equation} \label{eq:powerlawBB}
S_{\nu}(\lambda) = A_{BB} \frac{\lambda^{-(\beta+3) }}{e^{hc/\lambda k T}-1} + A_{PL} \lambda^{\alpha} e^{-(\lambda/\lambda_0)^2}.
\end{equation}

$S_{\nu}(\lambda)$ is in units of Jy, and $\alpha$ is the mid-IR power law index. Since the power-law dominates in the mid-IR and the blackbody component dominates at longer wavelengths, $\lambda_0$ is the wavelength at which the power law turns over and exponentially attenuates. 

\begin{figure}
\begin{center}
\includegraphics[width=1.0\columnwidth]{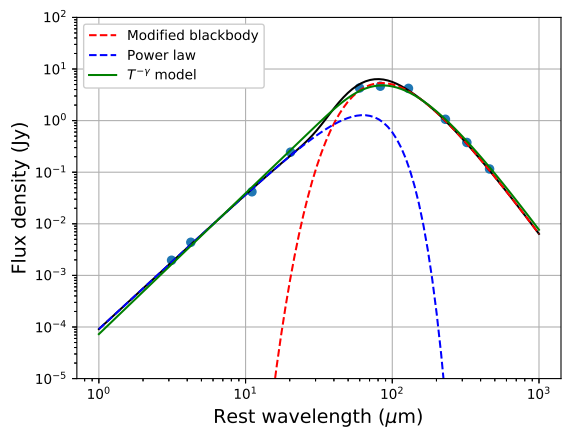}
\caption{{\label{fig:BB-PLsed} An example of an SED with the far-IR emission fit using a single temperature blackbody (red dashed curve) and a mid-IR power law (blue-dashed curve) that turns over and exponentially attenuates when the blackbody component dominates. The green curve shows the model in Eq. \ref{eq:tpowerlaw}.
}}
\end{center}
\end{figure}

Most of our SEDs have a limited number of far-IR bands available for fitting, so we need to reduce the number of free parameters in Eq. \ref{eq:powerlawBB}. That is, an SED with $N$ photometry bands allows for $N-1$ degrees of freedom in the fit. We first fit the mid-IR portion between $3<\lambda<50$ $\mu$m with a power law to obtain values for $A_{PL}$ and $\alpha$. A number of papers assume an emissivity parameter of $\beta=1.5$, which is empirically derived using from fitting modified blackbody functions to dusty emission \citep{Gordon_2010}. Here, we assume that $\beta=2$, which is the limiting value expected for dust grains in the Rayleigh regime, as constrained by the Kramers-Kronig relation \citep{Li_2005}. The precise value of $\beta$ does not significantly affect the total integrated IR luminosity or the dust temperature $T$ \citep{Casey_2012}. However, a larger value of $\beta$ increases the Rayleigh-Jeans portion of the far-IR emission such that it is in better agreement with the measured fluxes, and makes it unnecessary to add an additional cold dust temperature component. Fixing these values in Eq. \ref{eq:powerlawBB}, we fit the rest of the parameters to the SED. To ensure that the two components of Eq. \ref{eq:powerlawBB} transition smoothly, we fit the blackbody and power law components independently of each other to the photometry points with $\lambda>40$ $\mu$m and $\lambda<40$ $\mu$m, respectively. We then set $\lambda_0$ to be wavelength at which slopes are equal: $d\log B(\lambda)/d\log\lambda = \alpha$. Figure \ref{fig:BB-PLsed} shows an example of of an SED fit done using Eq. \ref{eq:powerlawBB}.

We require that there be at least three FIR band measurements to perform fitting of the FIR peak. To avoid any possible biasing in peak fitting and temperature measurements, we require that there be photometry measurements on both sides of the FIR peak for the fit to be considered reliable. Since there is a partial degeneracy between $\beta$ and $T$, we set $\beta=2$ to be a fixed parameter. There is a complete degeneracy between $T$ and $z$, so it is critical that we have accurate redshift values in order to accurately estimate the dust temperature. As discussed in Section \ref{section:photoz}, the $z_{photo}$ errors are sufficiently small that this will not significantly affect our dust temperature measurements. An error of $\sigma_z \sim 0.01$ produces an error of $\sigma_{\lambda} \sim 1$ $\mu$m for a typical far-IR peak wavelength of $\sim 100$ $\mu$m. This corresponds to an error of $\sigma_T \sim 0.3$K in Wien's displacement law, which is typically comparable or an order of magnitude smaller than the errors due to the modified blackbody fit itself. Our final sample contains 54 objects that have adequate photometry for far-IR dust fitting (see Figure \ref{fig:selection_flowchart}). This small sample size is likely the result of extreme Malmquist bias, since our requirements select the most IR luminous galaxies that are detectable with AKARI and Herschel SPIRE photometry. Since the majority of $L_{bol}$ comes from the far-IR dust peak of the SED, high levels of dust reddening might suggest that our sample is more heavily reddened than the usual population of dusty AGNs.

\begin{figure}
\begin{center}
\includegraphics[width=1.0\columnwidth]{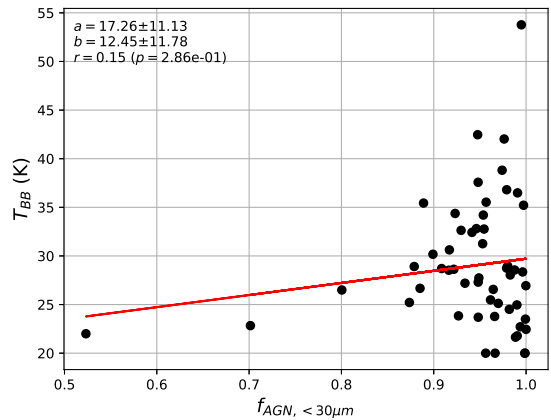}
\caption{{\label{fig:agnfrac-T} Correlation between AGN fraction $f_{AGN,<30\mu m}$ and the blackbody temperature $T_{BB}$ in Eq. \ref{eq:powerlawBB}. Red line shows the best fit.
}}
\end{center}
\end{figure}

Since different papers tend to use different values of $\beta$ (or even allowing $\beta$ to be a free parameter), it can be difficult to compare the values of $T_{BB}$ obtained from different studies with our results. However, we find that  value of $\beta$ does not significantly impact the value of $T_{BB}$, and only changes the slope of the long-wavelength portion of the fit. \citet{Riguccini_2015} fits a single temperature model to far-IR dust emission from a sample of DOGs in the COSMOS field, and finds a median temperature of $T \sim 40$K with $\beta=1.5$. This is higher than the median temperature of our sample ($T_{BB} \sim 28$K), but the range of temperatures generally agrees. If the central AGN engine is indeed responsible for heating the  dust, then we would expect the blackbody dust temperature to increase with the AGN fraction. As shown in Figure \ref{fig:agnfrac-T}, we do not find any significant trend between the AGN fraction $f_{AGN,<30\mu m}$ and the blackbody temperature $T_{BB}$ in Eq. \ref{eq:powerlawBB}. It is possible that a trend exists, but our small sample size prevents us from seeing this since we are selecting a smaller subset of objects from our SED analysis sample. The sample of DOGs in \citet{Riguccini_2015} contains a wide range of AGN fractions, and the authors shows that the AGN fraction is positively correlated with the dust temperature using a single temperature model.  Unfortunately the lack of objects with lower AGN fractions ($f_{AGN,<30\mu m} <0.8$) in our sample makes it difficult to determine if there is such a trend.

\subsection{Temperature power law distribution}
A more physical way to characterize dust emission is by using a temperature distribution, since there is no single characteristic $T$ that adequately describes the emission across the entire IR wavelength range. \citet{Kov_cs_2010} uses a power distribution of temperatures to  ($dM_d/dT \propto T^{-\gamma}$) parameterize the mid-IR emission, which in effect is very similar to using a power law parameterization of the flux. This temperature distribution $dM_d/dT$ is used to weight the modified blackbody function $\nu^{\beta} B_{\nu} (T)$. The total flux $S_{\nu}$ is given by:

\begin{equation} \label{eq:tpowerlaw}
S_{\nu}(\gamma,T_c) = (\gamma-1)T_c^{\gamma-1} \int ^{\infty} _{T_c} \nu^{\beta} B_{\nu} (T) T^{-\gamma} dT.
\end{equation}

\begin{figure}
\begin{center}
\includegraphics[width=1.0\columnwidth]{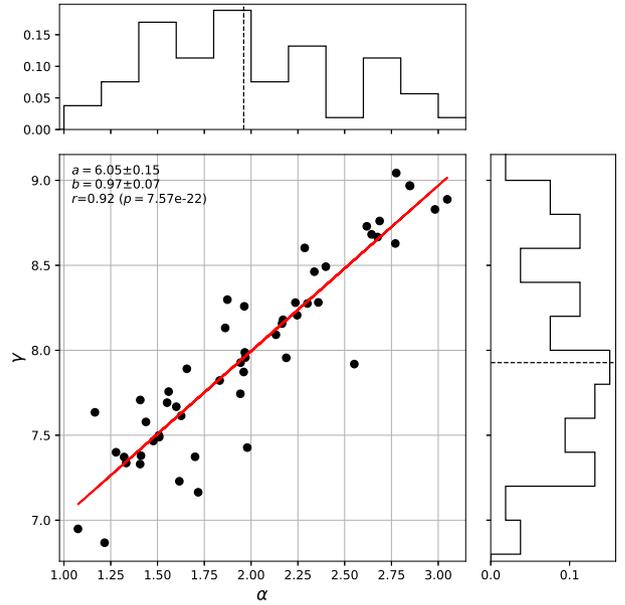}
\caption{{\label{fig:alpha-gamma} Correlation between $\alpha$ (mid-IR continuum power law index in Eq. \ref{eq:powerlawBB}) and $\gamma$ (temperature distribution power law index in Eq. \ref{eq:tpowerlaw}). Red line shows the best fit.
}}
\end{center}
\end{figure}

\begin{figure}
\begin{center}
\includegraphics[width=1.0\columnwidth]{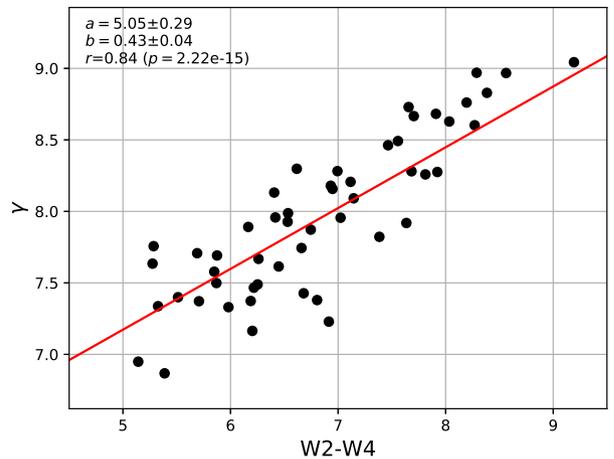}
\caption{{\label{fig:gamma-w2w4} Correlation between $\gamma$ (temperature distribution power law index in Eq. \ref{eq:tpowerlaw}) and $W2-W4$ WISE color. Red line shows the best fit.
}}
\end{center}
\end{figure}

This relation provides another two-parameter model to fit the dust emission. $\gamma$ determines the relative contribution of the hot dust temperature components, and thus determines the shape of the mid-IR emission. Since the mid-IR can be represented as a combination of multiple temperature components (modified blackbody functions) of decreasing amplitude, $T$ is well approximated by a power law distribution. For the purpose of numerical integration, we take $<10,000$K as the upper limit for the integral in Eq. \ref{eq:tpowerlaw}, since we are only concerned with the mid-IR emission. $T_c$ represents the coldest temperature dust component and this parameter governs the shape of the far-IR Rayleigh tail of the dust emission. Since both $\gamma$ and $\alpha$ parameterize the mid-IR continuum, Figure \ref{fig:alpha-gamma}  shows that the two are tightly correlated. Figure \ref{fig:gamma-w2w4} shows that $\gamma$ is also tightly correlated with WISE colors, and therefore this allows us parameterize the temperature distribution of hot dust using only WISE $W2-W4$ colors. This relation is not necessarily trivial, as $\gamma$ depends on a combination of both AKARI and WISE photometry.

The top histogram in Figure \ref{fig:alpha-gamma} shows the distribution of the mid-IR power law index $\alpha$. The tight relation between $\alpha$ and $\gamma$ shows that the mid-IR photometry can be used to constrain the radial density profile of the warm, compact dust producing the mid-IR emission. \citet{Blain_2002} measures range of $1.7< \alpha <2.2$ from a sample of dusty IRAS galaxies. Our measurements are centered around the same interval, but span a wider range ($1< \alpha <3$). Smaller values of $\alpha$ (i.e. flatter spectrum) are consistent with lots of warm dust, while a steep slope indicates very little dust.

The temperature distribution depends on the geometry and distribution of the emitting dust. $6.5 < \gamma < 
7.5$ corresponds to emission from sources within a diffuse medium, while $4 < \gamma < 5$ corresponds to emission from a dense medium at a distance from a heating source. The side histogram in Figure \ref{fig:alpha-gamma} shows that $\gamma$ has distribution with a median of $\langle \gamma \rangle \sim 8$, which indicates a small warm dust fraction and spatially diffuse dust distributed at farther distances from the central engine. This is not necessarily inconsistent with the compact torus model of AGN emission, since a compact dust torus can be embedded within a spatially larger, more diffuse dust mass distribution. In other words, the hot dust torus near the central AGN is heated to high temperatures (producing the mid-IR continuum), but there is also significant emission from dust at larger radii that represents the bulk of the dust mass. 

\begin{figure}
\begin{center}
\includegraphics[width=1.0\columnwidth]{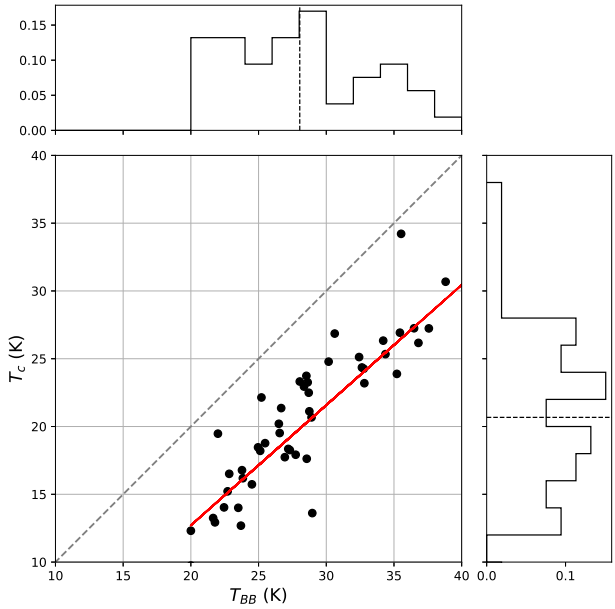}
\caption{{\label{fig:TBB-Tc} Correlation between $T_{BB}$ (modified blackbody temperature in Eq. \ref{eq:powerlawBB}) and $T_c$ (cold cutoff temperature in Eq. \ref{eq:tpowerlaw}). The grey dashed line indicates where $T_{BB} = T_c$. Red line shows the best fit.
}}
\end{center}
\end{figure}

Figure \ref{fig:TBB-Tc} shows that $T_{BB}$ and $T_c$ also have a tight linear correlation, and that $T_c$ is systematically lower than $T_{BB}$ by $\sim 8 $-$10$K. This shows than even in the Rayleigh tail of the dust peak, multiple temperature components are required to capture the shape of the far-IR emission. Even though $T_{BB}$ corresponds to the blackbody component centered around the far-IR peak and represents most of the far-IR emission, it does not represent the bulk of the dust \textit{mass} in the $dM/dT \propto T^{-\gamma}$ model. Therefore, dust mass measurements using an isothermal blackbody model can potentially lead to underestimates if the coldest temperature dust components are not taken into account.

\subsection{Correlation between AKARI fluxes and $T_{BB}$}
\begin{figure}
\begin{center}
\includegraphics[width=1.0\columnwidth]{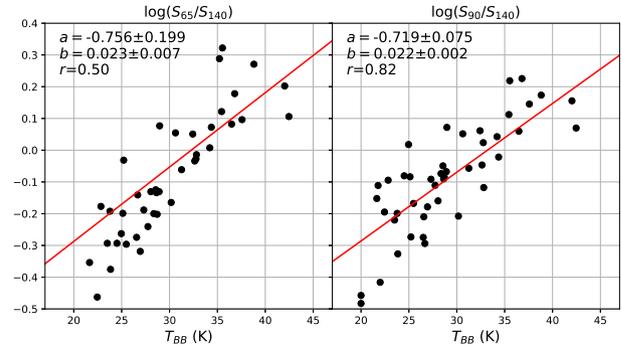}
\caption{{\label{fig:akariflux-Tbb} Correlation between AKARI flux ratios vs. the modified blackbody temperature in Eq. \ref{eq:powerlaw-BB}. Red line shows the best fit.
}}
\end{center}
\end{figure}

Since AKARI photometry samples the far-IR peak of the SED, we can use AKARI fluxes to predict the blackbody dust temperature based on the relative flux densities in different bands. This is potentially useful for estimating the dust temperature for galaxies that have a known redshift, but lack sufficient far-IR photometry to fit an emission model to its SED. Figure \ref{fig:akariflux-Tbb}  shows that the AKARI flux density ratios $\log (S_{65}/S_{140})$ and $\log (S_{90}/S_{140})$ are correlated with the blackbody temperature $T_{BB}$ in Eq. \ref{eq:powerlaw-BB}. The linear fits are given in the form:

\begin{equation}
\label{eq:Tbb-estimation}
T_{BB} = a + b \log \left ( \frac{S_{X}}{S_{Y}} \right ),
\end{equation}

where $X$ and $Y$ correspond to different wavelength AKARI bands. The correlation is weaker for $\log (S_{65}/S_{140})$, since the $S_{65}$ $\mu$m flux tends to be dominated by the mid-IR power law component of the dust model. The median power law turnover wavelength occurs at $\langle \lambda_c \rangle \sim 75$ $\mu$m (representing the transition between the warm and cold dust components), so $S_{90}$ is a better probe of the blackbody emission. There is also a positively correlated trend between these flux ratios and the cutoff temperature $T_c$ in Eq. \ref{eq:tpowerlaw}, but the scatter is large enough such that a linear regression fit is not useful for predicting $T_c$ to any reasonable accuracy. This is not surprising, since $T_c$ parameterizes the far-IR emission far beyond the wavelength range of AKARI. Even if AKARI flux ratios are not useful for predicting $T_c$, the photometry can be fit directly with Eq. \ref{eq:tpowerlaw} to place constraints on the parameters even if longer wavelength photometry (ex. from Herschel SPIRE) is not available. 

It's important to note that our AGN sample spans a relatively narrow range of redshifts ($z<0.15$), so the tightness of the correlation and fit parameters will likely change for a different redshift range. At higher $z$, this correlation will be less robust and perhaps disappear entirely since the AKARI bands will be sampling the mid-IR power law dominated region of the SED. This corresponds to $z > 0.9$, at which $S_{160}$ falls below the characteristic power law turnover wavelength $\lambda_c \sim 75$ $\mu$m.

\begin{figure}
\begin{center}
\includegraphics[width=1.0\columnwidth]{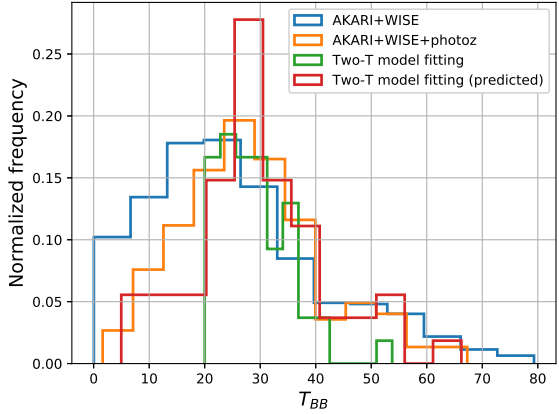}
\caption{{\label{fig:Tbb_estimation} Histograms of $T_{BB}$ values as estimated from Eq. \ref{eq:Tbb-estimation} and the relations in Figure \ref{fig:akariflux-Tbb}. The orange and blue curves show the temperatures for the AKARI-WISE AGN sample ($W1-W2>0.8$) with and without 2MASS $z_{photo}$ values, respectively. The green curve represents the temperatures fit directly to the SED with Eq. \ref{eq:powerlawBB}, and the red curve shows the values computed from Eq. \ref{eq:Tbb-estimation} for the same objects.
}}
\end{center}
\end{figure}

There are $N \sim 2,500$ objects in our original sample of WISE-selected AGN candidates with $W1-W2>0.8$, while only $N=56$ objects ($\sim 2\%$) have adequate far-IR photometry for direct temperature model fitting. We can use the relations in Figure \ref{fig:akariflux-Tbb} to estimate $T_{BB}$ for the remainder of these objects. The two regression fits have nearly identical fit parameters, but we choose to use $S_{90}/S_{140}$ since it has less scatter, and $S_{90}$ is less likely to fall into the mid-IR power law region at higher redshifts compared to $S_{65}$. Figure \ref{fig:Tbb_estimation} shows a comparison of the $T_{BB}$ values derived from $S_{90}/S_{140}$ for different subsets of our AGN sample. All of the distributions roughly coincide at the same $T_{BB}$ peak. In our temperature fitting procedure, there tends to be a lack of objects with $T_{BB}<20$ since there is less available photometry beyond the far-IR peak for most objects (likely biasing the temperature upwards). The larger WISE+AKARI sample lacking $z_{photo}$ measurements peaks at a slightly lower $T_{BB}$. The primary difference between this sample and the 2MASS photometric catalog sample (with $\langle z \rangle \sim 0.1$) is that the former presumably represents objects from a higher, wider range of redshifts (i.e. beyond $z>0.1$). At higher redshifts, the AKARI bandpasses are shifted in a way such that $S_{90}/S_{140}$ measurements are biased towards lower temperatures.

\subsection{IR luminosity} \label{section:IRluminosity}
Using the SEDs we computed in Section \ref{section:SEDmodeling}, we examine the relationships between $L_{IR}$ and other dust parameters. We adopt the convention in \citet{Kennicutt_Jr__1998} and define the IR luminosity to be between 8-1000 $\mu$m. Since this wavelength range represents around 70-80\% of $L_{bol}$, most of the trends applicable to $L_{IR}$ also holds for $L_{bol}$.

\begin{figure}
\begin{center}
\includegraphics[width=0.85\columnwidth]{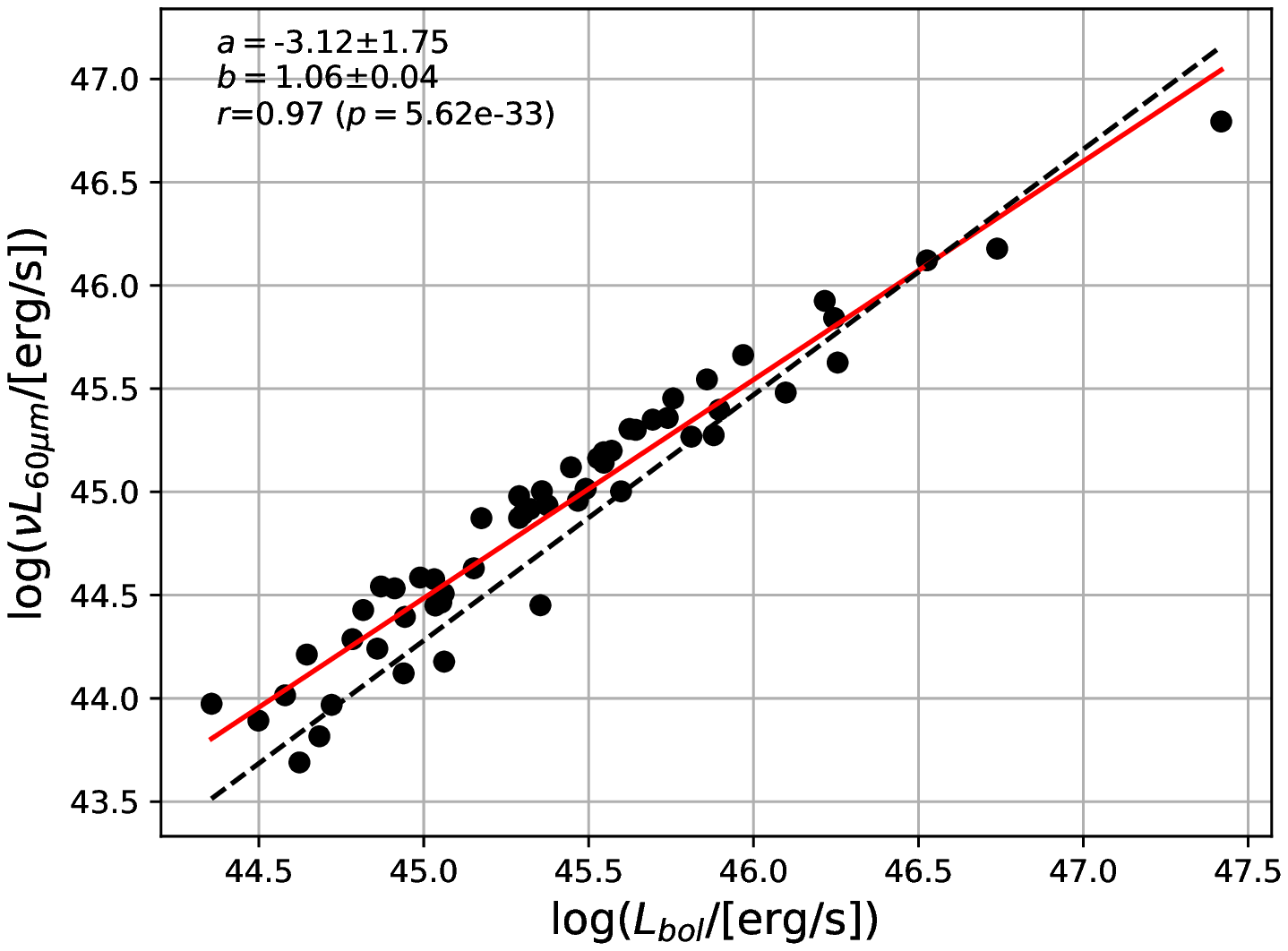}
\includegraphics[width=0.85\columnwidth]{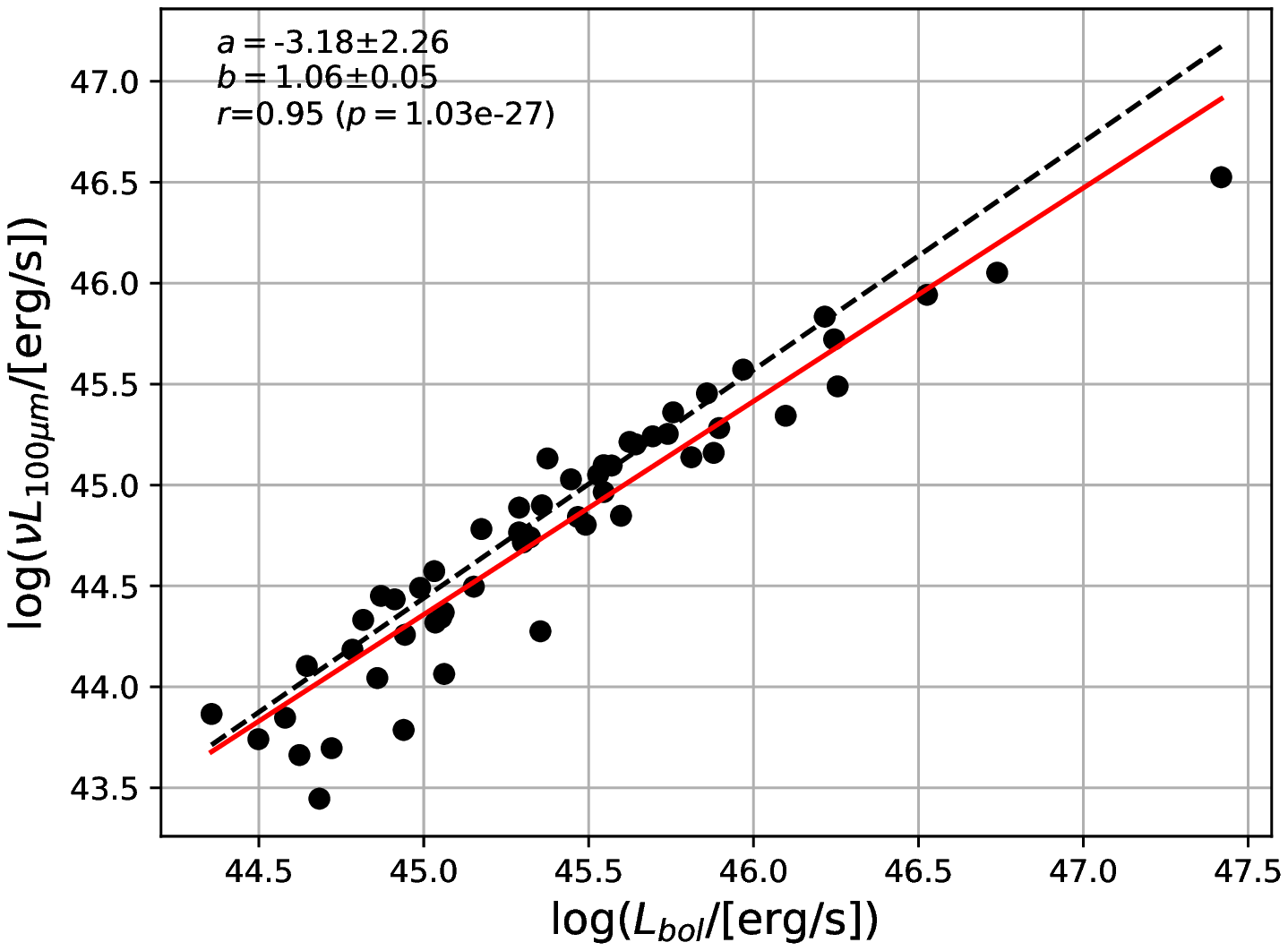}
\caption{{\label{fig:L_60um-L_bol} $\log (\nu L_{60 \mu m})$ and $\log (\nu L_{100 \mu m})$ plotted against $L_{bol}$. Red lines show the best fit, and the dashed lines are the relations derived in \citet{Spinoglio_1995}.
}}
\end{center}
\end{figure}

Figure \ref{fig:L_60um-L_bol} shows that there is a tight correlation between $L_{bol}$ and $L_{60\mu m}$ (as well as $L_{100\mu m}$). If we assume that the far-IR portion of the SED across different objects has a consistent shape and is well modeled by the same template, then this relation is trivial. The scatter in the correlation arises from the UV to mid-IR emission that is not included in the far-IR template. Figure \ref{fig:L_60um-L_bol} also shows the fits derived by \citet{Spinoglio_1995} from a sample of 12$\mu$m selected IRAS galaxies, which are in excellent agreement with our fits. The empirical data in \citet{Spinoglio_1995} show a larger scatter, so some caution is required when applying these luminosity correlations to individual objects. Since all of our AGNs are generally well fit with the far-IR templates of \citet{Mullaney_2011}, it is possible that this consistency in far-IR spectral shape is a result of our sample being limited to a narrow range of redshifts and luminosities (i.e. Malmquist bias), so we need to examine a more diverse set of objects to determine to whether such a relation holds more generally.

\begin{figure}
\begin{center}
\includegraphics[width=1.0\columnwidth]{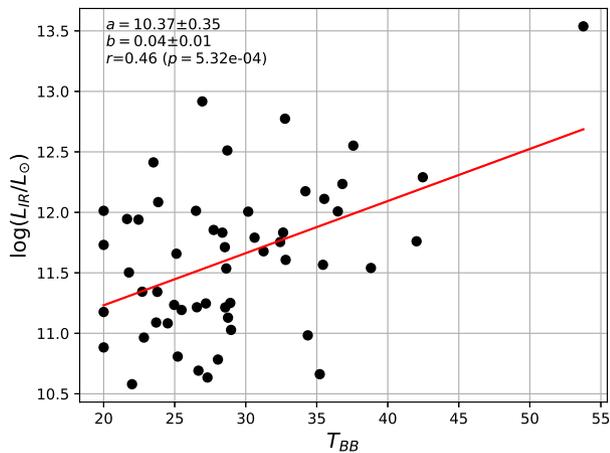}
\caption{{\label{fig:Tbb-L_IR} $\log(L_{IR})$ plotted against the blackbody temperature $T_{BB}$ in Eq. \ref{eq:powerlawBB}. Red line shows the best fit.
}}
\end{center}
\end{figure}

Even though dust temperatures are best modeled by a distribution instead of a single blackbody, a single temperature cold dust component captures the bulk of the emission in the far-IR. We would then expect there to be some sort of relation between this cold dust temperature and $L_{IR}$, since the bolometric luminosity of a blackbody increases monotonically with $T$. Figure \ref{fig:Tbb-L_IR} shows the dust temperature $T_{BB}$ (from Eq. \ref{eq:powerlaw-BB}) increases with increasing IR luminosity, consistent with trends found by \citet{Hwang_2010}, \citet{Elbaz_2010}, \citet{Amblard_2010}. However, our linear regression parameters do not agree with those obtained by other studies. We stress that the meaning of a dust temperature such as $T_{BB}$ is highly dependent on the model being used (ex. number of temperature components, availability of photometry, etc) and number of photometric bands available for fitting, so comparisons with the results of other studies are difficult.

\begin{figure}
\begin{center}
\includegraphics[width=1.0\columnwidth]{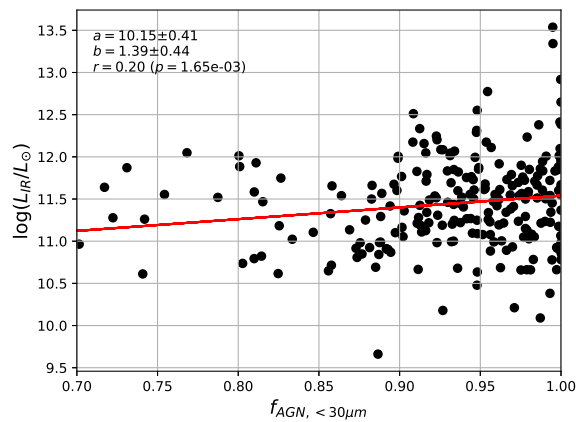}
\caption{{\label{fig:agnfrac-LIR} Fractional contribution of the AGN template to the total integrated SED luminosity at $<30\mu$m, plotted against the IR luminosity. Red line shows the best fit.
}}
\end{center}
\end{figure}

Figure \ref{fig:agnfrac-LIR} shows that there is a statistically significant positive correlation between the AGN fraction $f_{AGN,<30\mu m}$ and the IR luminosity, which is consistent with the idea that the dust emitting in the IR is being heated by the AGN engine.  \citet{Hwang_2013} also finds the same behavior for DOGs, and \citet{Veilleux_2009, Lee_2012,Ichikawa_2014, Riguccini_2015} find that in general AGN contribution increases with increasing $L_{IR}$. te

\subsection{Warm dust mass fractions} \label{subsection:warmdustmass}
We can integrate the power law temperature distribution $dM_d/dT \propto T^{-\gamma}$ to compute the dust mass:

\begin{equation} \label{eqn:dustmass}
M_d \propto \frac{A_d}{\kappa}\int ^{\infty} _{T_c} T^{-\gamma} dT,
\end{equation}

where $A_d$ is the effective radiative dust area and $\kappa$ is the effective dust opacity. Computing absolute values for $M_d$ is difficult because values of $\kappa$ for dust tend to be highly uncertain and dependent on wavelength. There are various parameterizations for $\kappa_{\nu}$ \citep{Scoville_1976}, but a detailed discussion of this is beyond the scope of this paper. Since we are primarily interested in seeing how the dust mass is dependent on AGN parameters and observables, we instead compute dust mass ratios, which only depend on the SED fit parameters $T_c$ and $\gamma$. This allows for easy comparisons between individual objects and AGN populations without any assumptions regarding dust opacity models. However, this does assume that the effective dust area $A_d$ is relatively constant across objects.

Dust mass estimates that rely on a single temperature blackbody component tend to be highly sensitive to the value of $T_{BB}$ \citep{Draine_2007}. This is even more so in the case of using a temperature power law distribution, in which the cold temperature dust masses increase without bound with $T_c$. Since the total mass scales with $T_c$ as $M_d \propto 1/T_c^{-\gamma+1}$, a typical value of $\gamma \sim 8$ causes the dust mass to be highly sensitive to the measured value of $T_c$. This extreme dependence makes computing the total dust mass difficult, since a small uncertainty in $T_c$ can change the mass estimate by a few orders of magnitude. Additionally, it is difficult to precisely define what the total dust mass content actually means, since the far-IR SED emission represents the dust that is heated and detectable, and does not include any colder dust that does not contribute to the emission.

Rather than estimating the total dust mass at all temperatures, we instead compute the relative fraction of warm and cold dust components given fixed temperature ranges. This leaves $\gamma$ as a free parameter, which has relatively low uncertainties since the mid-IR continuum is well sampled and well fit by a power law. If we assume that each temperature corresponds to a different radius from the central AGN, knowing the relative masses of the different temperature components provides information about the radial mass distribution.

We define the warm dust mass to be $T>50$K, and the total dust emission to be all the mass with $T>15$K. Since the integral in Eq. \ref{eqn:dustmass} is just a simple power law, the distribution only depends on $\gamma$. The choice of temperature bounds is somewhat arbitrary, since the bounds do not affect the distribution of dust mass fractions and only shifts the fractions by a constant multiplicative factor.  

\begin{figure}
\begin{center}
\includegraphics[width=1.0\columnwidth]{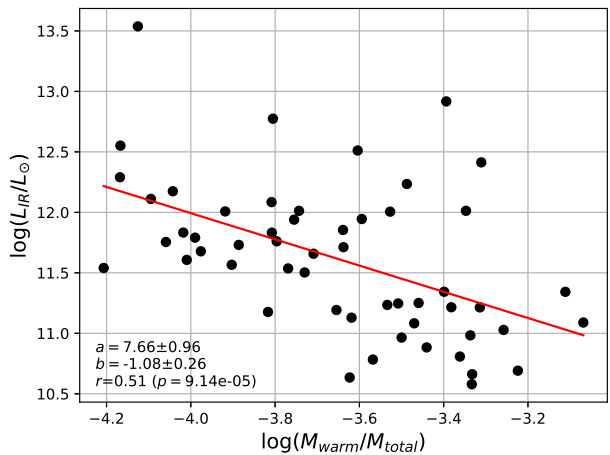}
\caption{{\label{fig:dustmass-L_IR} $L_{IR}/L_{\odot}$ plotted against the warm dust fraction $\log(M_{warm}/M_{total})$. Red line shows the best fit.
}}
\end{center}
\end{figure}

Figure \ref{fig:dustmass-L_IR} shows that $L_{IR}$ is negatively correlated with the warm dust mass fraction. In other words, this implies that the cold dust mass fraction is \textit{positively} correlated with $L_{IR}$. This supports the idea that the bulk of the IR luminosity is dominated by cold dust due to its significantly larger mass compared to warm dust. We also observe no correlation between the optical extinction $E(B-V)$ and dust mass. As discussed earlier, this is not too surprising given that the extinction likely depends on factors (ex. geometry, covering fraction, etc.) other than the total dust content. Since we have a very narrow range of AGN fractions ($f_{AGN<30 \mu m}$), we were unable to find any significant relationship between warm mass dust fraction and AGN fraction. Such a positive correlation would give support to the idea that the AGN is responsible for heating the warm dust located closer to the central engine. As was the case for AGN fraction and $T_{BB}$ (see Figure \ref{fig:agnfrac-T}), we would need a sample of AGNs with a wider range of AGN fractions to demonstrate this relation.

\section{Conclusion} \label{conclusion}
Using only archival photometric data and simple SED models, we have demonstrated that we can identify dust-obscured AGNs and understand their physical properties. SED analysis allows us to efficiently identify obscured AGNs without the need for time-intensive spectroscopy. Even without far-IR photometry, we have shown that mid-IR data from WISE can be used to estimate bolometric luminosities and using simple analytical models with only 3-4 parameters, we can quantify dust temperature, mass, and spatial distribution using only SED analysis. This approach greatly increases the number of AGNs we can analyze given the abundance of photometry relative to spectroscopic data, without the time and expense of obtaining the latter.

The majority of our findings regarding the nature of dusty AGNs are consistent with generally accepted AGN models. Objects that are very red in the mid-IR are dominated by an AGN component (i.e. power law behavior), and this emission due to warm dust (which is only a few percent of the total dust mass) represents a significant fraction of the bolometric flux. We find that AGNs generally have the same shape in their far-IR SEDs across different objects, which validates one of the assumptions in applying bolometric corrections to mid-IR luminosities. We have shown trends suggesting that the central AGN is responsible for heating its surrounding dust, but to demonstrate this more definitively we need to examine a sample of AGNs that have a wider range of AGN fractions. The spatial dust distribution that we infer from our dust temperature models shows that the dust is diffuse and extended, which is somewhat at odds with the standard AGN compact torus model. To examine this further, we would need to examine resolved imaging of these objects.

For future work, we should conduct this analysis on a sample of AGNs with a wider range of redshifts to see if our results hold for galaxies located beyond those in the nearby Universe. One question that remains is whether the dust temperature or warm dust fraction systematically changes with redshift. This analysis should also be performed on a sample of less luminous galaxies that we were not able to examine due to Malmquist-biased nature of our catalog. Future surveys and deeper redshift catalogs (ex. all-sky spectroscopic redshifts from the SPHEREx mission) will allow us to answer these questions.

\section*{Acknowledgements}
This research is based on observations with AKARI, a JAXA project with the participation of ESA. This publication makes use of data products from the Wide-field Infrared Survey Explorer, which is a joint project of the University of California, Los Angeles, and the Jet Propulsion Laboratory/California Institute of Technology, funded by the National Aeronautics and Space Administration. Herschel is an ESA space observatory with science instruments provided by European-led Principal Investigator consortia and with important participation from NASA. The Pan-STARRS1 Surveys (PS1) and the PS1 public science archive have been made possible through contributions by the Institute for Astronomy, the University of Hawaii, the Pan-STARRS Project Office, the Max-Planck Society and its participating institutes, the Max Planck Institute for Astronomy, Heidelberg and the Max Planck Institute for Extraterrestrial Physics, Garching, The Johns Hopkins University, Durham University, the University of Edinburgh, the Queen's University Belfast, the Harvard-Smithsonian Center for Astrophysics, the Las Cumbres Observatory Global Telescope Network Incorporated, the National Central University of Taiwan, the Space Telescope Science Institute, the National Aeronautics and Space Administration under Grant No. NNX08AR22G issued through the Planetary Science Division of the NASA Science Mission Directorate, the National Science Foundation Grant No. AST-1238877, the University of Maryland, Eotvos Lorand University (ELTE), the Los Alamos National Laboratory, and the Gordon and Betty Moore Foundation. This research has made use of the SVO Filter Profile Service (http://svo2.cab.inta-csic.es/theory/fps/) supported from the Spanish MINECO through grant AyA2014-55216.


\end{document}